# Modeling of parallel power MOSFETs in steady-state

Minh Nhat Huynh[1, *], Minh Khoi Nguyen Tien[1], Cong Toai Truong[1,2,3], Minh Tri Nguyen[1], Quoc Minh Lam[1], Van Tu Duong[1,2,3], Huy Hung Nguyen[1,4], Tan Tien Nguyen[1,2,3, †]

[1] National Key Laboratory of Digital Control and System Engineering (DCSELab), Ho Chi Minh City University of Technology (HCMUT), 268 Ly Thuong Kiet Street, District 10, Ho Chi Minh City 700000, Vietnam
[2] Faculty of Mechanical Engineering, Ho Chi Minh City University of Technology (HCMUT), 268 Ly Thuong Kiet, District 10, Ho Chi Minh City 700000, Vietnam
[3] Vietnam National University Ho Chi Minh City, Linh Trung Ward, Thu Duc District, Ho Chi Minh City 700000, Vietnam
[4] Faculty of Electronics and Telecommunication, Saigon University, Ho Chi Minh City 700000, Vietnam

[†]Leading - Corresponding author: nttien@hcmut.edu.vn
[*]Presenting author: nhat.huynh1905758@hcmut.edu.vn

**Abstract**

In high–power applications, multiple power MOSFETs are connected in parallel and treated as a single switch in order to handle much larger total currents. In this paper, a parallel power MOSFETs model from the turn–off state until they reach their steady state is introduced. The model represents the relationship between each power MOSFET's gate voltage and the current distribution among them. The study's key purpose is to use the model for dealing with the asymmetry in sharing current and power loss between these semiconductor devices during the steady–state region.

## 1 Introduction

The need for renewable energy sources to replace traditional fossil fuels, as well as the increasing in demand for high–power applications such as electric vehicles (EVs), is driving the advancement of battery technology and its applications in the present day [1], [2]. The utilization of high–voltage batteries in high–power applications has elevated the requirement for protection. As the discharge current increases, the discharge process becomes more hazardous. Overcurrent protection is essential in maintaining the safety of the batteries during operation. There are various methods for interrupting DC currents such as mechanical breakers or semiconductor devices [3].

In case there is a malfunction in the system, a circuit for interrupt the power source from the system is necessary. The main switch for DC circuit breakers can be either contactors or a SiC MOSFETs. Contactors are appropriate for applications with high current loads because they can handle high currents. When they do, though, an arc might form inside the contactor. In the worst–case scenario, they may experience welding or bouncing which shortens the lifespan of the contactor as a consequence. Besides, for protection purposes, the time it takes to restore and eliminate a fault in a security system is one of the most important parameters. They must be quick enough to repair any potentially deadly harm done to battery cells and other electrical parts. Contactors are unable to quickly convert to their blocking state due to the presence of moving parts during the switching state [4]–[6].

On the other hand, a solid-state circuit breaker uses semiconductor devices in place of the conventional mechanical breaker. They can transition states quickly as a result of the lack of moving elements, making them ideal for applications that require quick protection from overcurrents or short circuits [7]. Additionally, the use of semiconductor devices eliminates the problem of arcs emitted during switching, making solid-state circuit breakers a more reliable solution.

Nevertheless, power MOSFETs are not typically designed for handling hundreds of amperes on their own. As a result, the concept of paralleling multiple MOSFETs is unavoidable to achieve the required current ratings without breaking individuals' operating limits [8]–[10]. This method also reduces the equivalent on–state resistance which leads to a reduction in power dissipation.

This way, the current and power are expected to distribute equally between them, reducing the burden on each individual. In reality, because of various reasons, it is a key challenge to ensure the current sharing among power MOSFETs, which leads to unevenly distributed power loss and thermal dissymmetry [11]. This paper focuses on the studying of uneven sharing current between paralleled SiC power MOSFETs.

## 2 Current imbalance

Paralleled power MOSFETs might experience unbalance in sharing current during two separate phases.

### 2.1. Dynamic current imbalance

One of them is dynamic current imbalance, which occurs during the MOSFETs' turn–on and turn–off process. Unbalanced current sharing throughout the state can cause high current spikes and overloading in certain MOSFETs, leading to reduced performance and potential thermal runaway. The imbalances stem from variations in switching time among the parallel power MOSFETs, which are affected by mismatches in threshold voltage ($V_{TH}$) and transconductance ($g_{fs}$). In addition to the device parameter mismatch, asymmetry in the cable design can also contribute to the uneven distribution of current [12], [13]. Even though unequal current sharing during the state can have a significant impact on high–frequency applications, it only affects the DC circuit breaker for a brief moment during a fault occurrence.

### 2.2. Static current imbalance

On the other hand, static current imbalance directly affects the protection device's principle because most of the time, the paralleled power MOSFETs act as a conduction device. During on–state, the power MOSFETs can be considered as equivalent resistors with the values equal to their on–state resistance $(R_{DS(on)})$. For SiC power MOSFETs, $R_{DS(on)}$ can obtain the value of hundreds or even tens of $m\Omega$ with the tolerance of $\pm 20\%$. These parallel MOSFETs all share the same drain–to–source voltage $(V_{DS})$, therefore the lower on–state resistance of a particular MOSFET, the higher the drain current $(I_D)$ flowing through it, resulting in increased power dissipation. The devices may also experience thermal stress due to uneven heating, which may result in permanent damage. Consequently, this might lead to reducing in reliability, device lifetime, and even device failure.

One way to mitigate the current unbalance in paralleled power MOSFETs is by relying on positive temperature coefficient (PTC) elements. A PTC element is a material that has a resistance that increases as its temperature rises, which can be used to suppress the uneven in sharing current among the parallel devices to some extend. During normal operation, the MOSFETs with lower $R_{DS(on)}$ will dissipate more power and heat, causing their PTC elements to have a higher resistance and reducing the drain current flowing through them as a result. This helps equalize the drain current distribution among the paralleled MOSFETs, preventing thermal stress, hot spots, and prolonging the device's lifespan, which is called thermal self–stabilization.

Although SiC MOSFETs' positive temperature coefficient can contribute to current sharing in some situations, it might not always be enough to guarantee balanced current distribution in high current applications. This is because when a large amount of current passes through the system, even a small imbalance in current distribution can result in significant power dissipation and temperature rise in one or more of the MOSFETs. Besides, the use of PTC elements to balance current in paralleled SiC MOSFETs may not always be effective if the current imbalance is too severe. The sensitivity of $R_{DS(on)}$ to temperature varies based on the type of MOSFET, as shown in the equation below [14]. Consequently, the higher temperatures can cause degradation of the MOSFETs over time, reducing their performance and lifespan. Therefore, relying exclusively on PTC for suppressing static current imbalance is inappropriate.

$$R_{DS(on)-t} = R_{DS(on)-25} \left(\frac{t + 273.15}{298.15}\right)^n \tag{1}$$

Where: $R_{DS(on)-t}$ is MOSFETs on–state resistance at temperature $t$ (°C); $R_{DS(on)-25}$ is MOSFETs on-state resistance at 25°C; and $n$ is the fitting parameter. This parameter varies depending on the type of MOSFET, which is particularly low in SiC devices compared to SI one [15].

### 2.3. Technique for active load balancing

In addition to being temperature-dependent, power MOSFETs are also voltage–driven devices, in which their conductivity capabilities vary in response to the gate-to-source voltage $(V_{GS})$ applied to them. Based on [16], the model for simulating the drain current $I_D$ of a SiC power MOSFET can be calculated by equations (2) and (3):

During linear region: $V_{GS} > V_{TH}; V_{DS} < V_{GS} - V_{TH}$

$$I_D = \mu_n C_{ox} \frac{W}{L} \left[ (V_{GS} - V_{TH}) V_{DS} - \frac{V_{DS}^2}{2} \right] \quad (2)$$

During saturated region: $V_{GS} > V_{TH}; V_{DS} \geq V_{GS} - V_{TH}$

$$I_D = \frac{1}{2} \mu_n C_{ox} \frac{W}{L} (V_{GS} - V_{TH})^2 \quad (3)$$

Where: $\mu_n$ is electron mobility; $C_{ox}$ is oxide capacitance; $W$ is width of the gate; $L$ is length of the gate; $V_{DS}$ is drain–to–source voltage; $V_{GS}$ is gate–to–source voltage; and $V_{TH}$ is gate threshold voltage.

As aforementioned, their resistance can obtain very small values of hundreds or even tens $m\Omega$, which leads to a much smaller $V_{DS}$ compare to the difference between gate–to–source voltage and threshold voltage $(V_{GS} - V_{TH})$. Therefore, based on [17] and equation (4), the drain current during on–state period can be determined as shown:

$$I_D = \mu_n C_{ox} \frac{W}{L} (V_{GS} - V_{TH}) V_{DS} \text{ for } (V_{DS} \ll (V_{GS} - V_{TH})) \quad (5)$$

As a result, the on–state resistance can be expressed as an equation depends on $V_{GS}$:

$$R_{DS(on)} = \frac{1}{\mu_n C_{ox} \frac{W}{L} (V_{GS} - V_{TH})} \quad (6)$$

Equation (5) demonstrates that the power MOSFET can be considered as a voltage control resistor within the steady–state region. In particular, $R_{DS(on)}$ can be varied depends on $V_{GS}$, which leads to a method for actively balance the sharing current.

## 3 Investigation of static current unbalance suppression

### 3.1. Overall system description

A system has been constructed to monitor the distribution of current among power MOSFETs when they are connected in parallel. Additionally, the system is designed to assess the effect of both temperature and $V_{GS}$ on the current sharing. This is achieved by allowing the system to observe the current flowing through each MOSFET as well as their temperature, and by enabling the adjustment of the $V_{GS}$ applied to each MOSFET separately. The system includes:

1. Two IRFP4688 MOSFETs.
2. Two ACS758 hall current sensors.
3. Two NTC Thermistor Sensors.
4. Dummy loader.

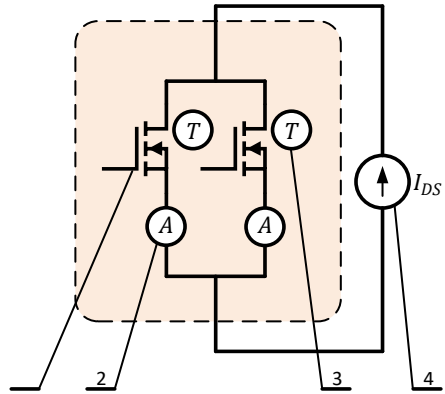

**Fig. 1.** Principle of steady–state paralleled MOSFETs

### 3.2. Experiment model

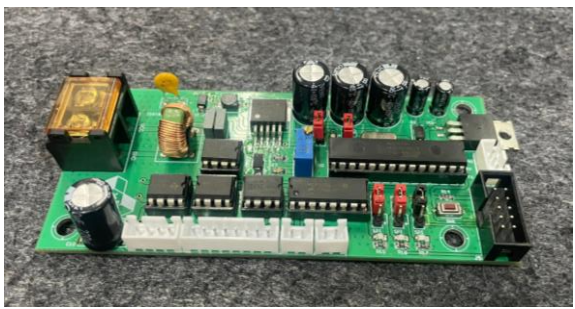 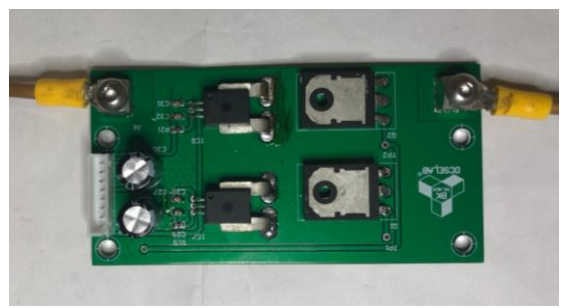

(a) Control circuit  (b) Power circuit

**Fig. 2.** Printed circuit board of steady – state paralleled MOSFETs

**Fig. 2** depicts a printed circuit board (PCB) that is designed to conduct test cases with parallel MOSFETs. **Fig. 2a** is the control board, featuring an AVR chip at its core that guarantees the consistency of pulse width modulation and data collection. **Fig. 2b** displays an integrated circuit of two IRFP4688 MOSFETs connected in parallel, similar to the system description outlined earlier. The division of control and dynamic circuits into separate entities helps to minimize interference and ensures the accuracy of the experimental results by avoiding any potential errors.

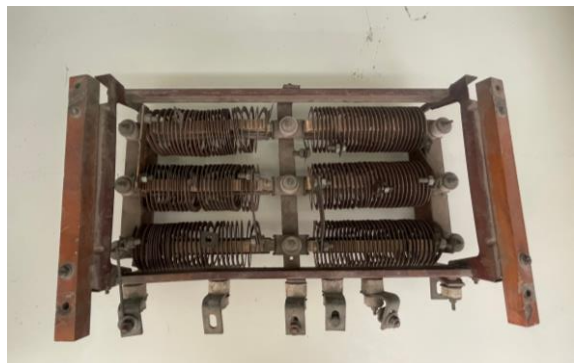

**Fig. 3.** Dummy loader for testing paralleled MOSFETs

**Fig. 3** shows the Dummy loader used in this paper. The dummy load system consists of six connected load resistors, with their specific parameters outlined in **Table 1**.

**Table 1.** The parameters of the dummy loader

| Parameter | Unit | Value |
|---|---|---|
| Overall size | $mm$ | $600 \times 400 \times 150 mm$ |
| Mass | $kg$ | 10 |
| Range | $\Omega$ | $[0 - 6]$ |
| Resolutions | $\Omega$ | $1/3$ |

### 3.3. Experiment and results

*Case 1: Using ARX Method to identification of one* MOSFET.

First, individual experiments were performed on each MOSFET to find an equation representing the characteristics of the IRFP4688 MOSFETs. The condition experiment is shown in **Table 2**.

**Table 2.** The parameters of the resistance load

| Parameter | Unit | Value |
|---|---|---|
| Power |  | LiFePO4 24V (20Ah – 5C) |
| Temperature | °C | 25 |
| Number of MOSFETs | pieces | 10 |
| Sampling time | s | 0.05 |

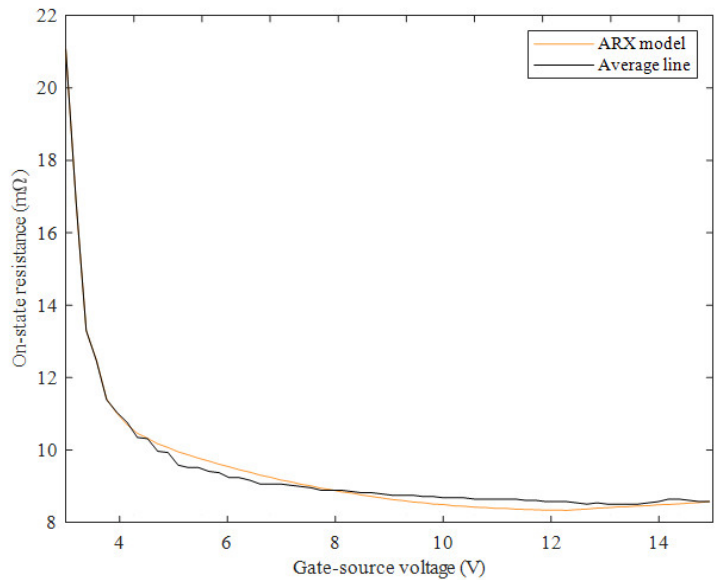

**Fig. 4.** Average data and ARX model of 10 MOSFETs

The relationship between $V_{GS}$, and $R_{DS}$, is modeled using the ARX model. **Fig. 4** displays the average data line of 10 MOSFETs (represented by the black line) and the ARX model (represented by the orange line) of the system. The obtained discrete–time ARX model is:

$$A(z)y(t) = B(z)u(t) + e(t)$$

With:

$$A(z) = 1 - 0.955z^{-1} - 0.336z^{-2} + 0.218z^{-3} + 0.056z^{-4}$$

$$B(z) = -0.005z^{-6} + 0.054z^{-7} - 0.054z^{-8} + 0.016z^{-9}$$

*Case 2: Test the model was founded with a different MOSFET from the previously obtained data set.*

With the same experimental conditions as case 1, the data of the 11th MOSFET is collected and shown in **Fig. 5**. The test data line has a degree of compatibility with the model (error of about 3%).

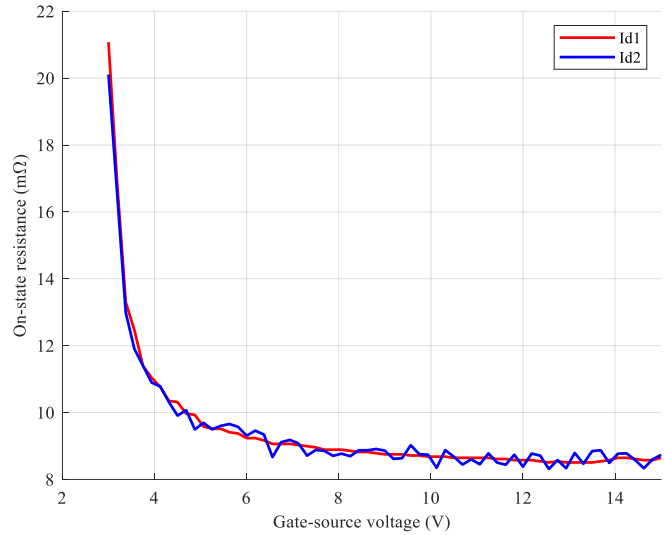

**Fig. 5.** Evaluate ARX model

*Case 3: The case of parallel MOSFETs connected.*

Experimental conditions of the parallel MOSFET model are shown in **Table 3**.

**Table 3.** The parameters of the resistance load

| Parameter | Unit | Value |
|---|---|---|
| Power |  | LiFePO4 24V (20Ah – 5C) |
| Temperature | °C | $25^o$ |
| Time of experiment | min | 20 |
| Sampling time | $ms$ | 50 |

The process of observing the thermal and current balance through pulse width modulation ($V_{GS}$) is difficult to accomplish when working with a single MOSFET. To overcome this challenge, the experiment was conducted with two paralleled MOSFETs so that changes in temperature and current per MOSFET could be clearly observed. The progress is presented below.

An experiment is conducted to evaluate the capability of thermal self–regulation by applying a total current of 20A to the system. During the experiment, each MOSFET is activated with the maximum $V_{GS}$ while recording both the temperature and the current flowing through each device.

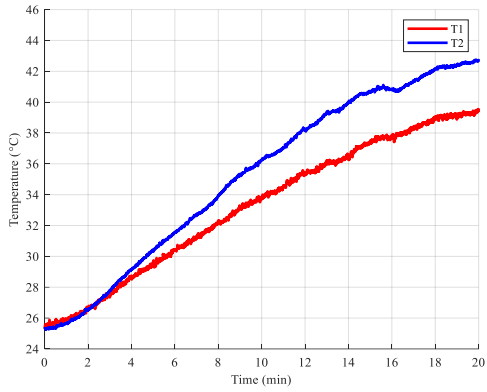 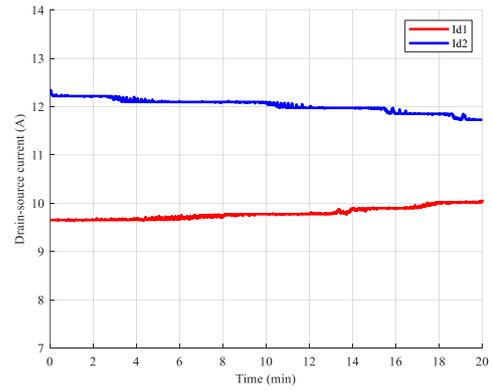

(a) Temperature of two MOSFETs
(b) Drain current of 2 MOSFETs

**Fig. 6.** Sharing drain current and thermal self–regulation of paralleled MOSFETs

The collected data reveals a constraint in relying solely on their PTC element to balance the discharge current, which is especially pronounced for SiC devices. The graph illustrates the temperature discrepancy between the two devices. Although the PTC element helps prevent the worst outcome – thermal runaway, it still negatively impacts the lifespan and reliability of the system, especially in high–power applications. As a result, in order to entirely suppress the uneven in sharing current, it is advisable to implement an active load balancing technique.

As aforementioned, MOSFETs' on–state resistance can be controled by adjusting $V_{GS}$ applied to each MOSFETs. By manipulating the gate–source voltage, it becomes feasible to regulate the current that flows through each MOSFET and achieve a balanced distribution of current among paralleled MOSFETs.

An experiment is performed to assess the viability of the method mentioned above. This involves applying the maximum gate voltage to one device and altering the gate voltage on the other device, while simultaneously monitoring the flow of current between two devices.

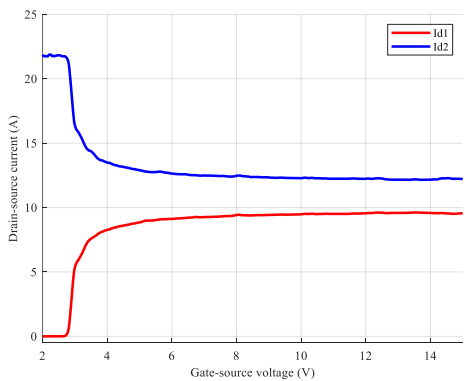 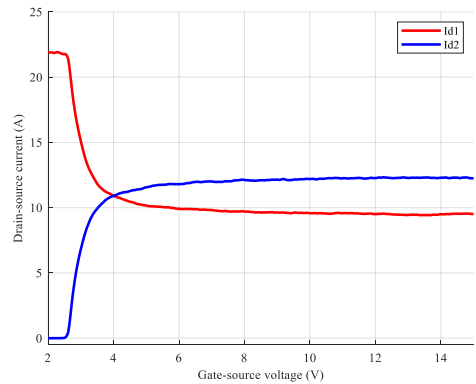

(a) Adjusting 1st MOSFET's gate voltage
(b) Adjusting 2nd MOSFET's gate voltage

**Fig. 7.** Adjust sharing current between two MOSFETs

The results of the experiment (show in **Fig. 7**) demonstrate that the division of current among the MOSFETs can be significantly altered. The figure highlights the presence of an intersection point where the shared current between the two MOSFETs reaches a state of balance. This outcome implies that the active balancing method, facilitated by the adjustment of $V_{GS}$, can maintain a balanced load current even under varying conditions, further enhancing the reliability and efficiency of the system.

## 4 Conclusions

For the purpose of improving safety and reliability in high–power applications, a DC circuit breaker that utilizes parallel MOSFETs as the primary switch is an appropriate solution. This research examines the issue of unequal current division among these paralleled MOSFETs during steady–state and offers a means to resolve it. The results reveal that a balanced state among MOSFETs can be achieved more effectively and safely through the adjustment of their $R_{DS(on)}$ values by means of active balancing in stead of solely rely on PTC element.

## Acknowledgements

This research is funded by Vietnam National University Ho Chi Minh City (VNU-HCM) under grant number TX2023-20b-01. We acknowledge the support of time and facilities from National Key Laboratory of Digital Control and System Engineering (DCSELab), Ho Chi Minh City University of Technology (HCMUT), VNU-HCM for this study.